\pdfoutput=1
\documentclass[11pt,a4paper,twoside,openright,closeany,textany]{book}
\usepackage{makeidx}
\usepackage{graphicx}
\usepackage{paralist}
\usepackage{caption} 
\usepackage{amsmath}
\usepackage{amssymb}
\usepackage{mathtools}
\usepackage{fancyhdr}
\usepackage{multicol}
\usepackage{multirow}

\usepackage{mdframed}
\usepackage[usenames,dvipsnames,svgnames,table]{xcolor}
\usepackage{array}
\usepackage[%
  colorlinks=true,%
  linkcolor=black,%
  urlcolor=black,%
  citecolor=black%
]{hyperref}
\usepackage{nameref} 
\usepackage{ifthen} 
\usepackage[font={small,sl}]{caption}
\usepackage[authoryear]{natbib}
\renewcommand\bibname{References}
\newcommand{\mychapbib}{
  \addcontentsline{toc}{section}{\bibname}
  \bibliographystyle{natbib}
  \bibliography{strucbioinf}
}
\usepackage[colorinlistoftodos]{todonotes}
\usepackage{letltxmacro}

\usepackage{tocstyle} 
\usetocstyle{standard}
\settocfeature{raggedhook}{\raggedright}
\newcommand{\bibfont}{\small\raggedright}
\def\cite{\citep}

\LetLtxMacro{\oldTodo}{\todo}
\renewcommand{\todo}[2][]{\oldTodo[#1]{TODO: #2}}

\usepackage[normalem]{ulem}

\newcommand\inwish[1]{\oldTodo[inline,color=SkyBlue]{WISH: #1}}

\newboolean{onechapter}

\newcommand{\AF}[1][~]{K.\@#1Anton#1Feenstra}
\newcommand{\SA}[1][~]{Sanne#1Abeln}

\newcommand{\HM}[1][~]{Halima#1Mouhib}

\newcommand{\JvG}[1][~]{Juami#1H.\@#1M.\@#1van#1Gils}

\newcommand{\MD}[1][~]{Maurits#1Dijkstra}

\newcommand{\JV}[1][~]{\mbox{Jocelyne}#1\mbox{Vreede}}

\newcommand{\AR}[1][~]{\mbox{Arri\"en}#1\mbox{Symon}#1\mbox{Rauh}}

\newcommand{\orcid}[1]{\href{https://orcid.org/#1}{\raisebox{-0.7ex}{\protect\includegraphics[height=3ex]{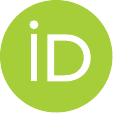}}}}
\definecolor{idgreen}{RGB}{166 206 57}
\newcommand{\mailid}[1]{\href{mailto:#1}{\raisebox{-0.3ex}{\color{idgreen}\textsf{\textbf{\Large \protect@}}}}}

\newcommand{\AFid}{\orcid{0000-0001-6755-9667}}
\newcommand{\SAid}{\orcid{0000-0002-2779-7174}}
\newcommand{\HMid}{\orcid{0000-0001-5031-3468}}

\newcommand{\JvGid}{\orcid{0000-0003-3706-7818}}

\newcommand{\JVid}{\orcid{0000-0002-6977-6603}}

\newcommand{\ARid}{\orcid{0000-0001-9707-3836}}

\newcommand{\MDid}{\orcid{0000-0002-7971-6209}}

\newcommand{\ACtxt}{Wrote the text}
\newcommand{\ACfig}{Created figures}
\newcommand{\ACref}{Review of current literature}
\newcommand{\ACeds}{Editorial responsibility}
\newcommand{\ACproof}{Critical proofreading}
\newcommand{\ACfb}{Non-expert feedback}

\newcommand{\Angs}[1][~]{\text{\normalfont\AA}}

\renewcommand{\and}{\quad}

\newcommand{\pdbref}[1]{\href{http://www.rcsb.org/pdb/explore.do?structureId=#1}{PDB:#1}}
\newcommand{\arxiv}[2][UNDEFINED]{\href{https://arxiv.org/abs/#2}{\ifthenelse{\equal{#1}{UNDEFINED}}{arxiv.org/abs/#2}{#1}}}

\newcommand{\figref}[2][]{\hyperref[fig:#2]{Figure\@~\ref*{fig:#2}#1}}
\newcommand{\tabref}[1]{\hyperref[tab:#1]{Table \ref*{tab:#1}}}
\renewcommand{\eqref}[2][]{\hyperref[eq:#2]{Equation#1\@~\ref*{eq:#2}}}
\newcommand{\panelref}[2][]{%
    \ifthenelse{\boolean{onechapter}}{%
        \hyperref[panel:#2]{Panel\@~``\nameref{panel:#2}#1''}%
    }{%
        \hyperref[panel:#2]{Panel\@~\ref*{panel:#2}#1}%
    }%
}
\newcommand{\secref}[2][n]{%
    \hyperref[sec:#2]{%
        \ifthenelse{\equal{#1}{n} }{Section\@~\ref*{sec:#2}}{}% just number
        \ifthenelse{\equal{#1}{nn}}{Section\@~\ref*{sec:#2} ``\nameref{sec:#2}''}{}% nm & nr
        \ifthenelse{\equal{#1}{N} }{``\nameref{sec:#2}''}{}% just quoted name
        \ifthenelse{\equal{#1}{NN} }{\nameref{sec:#2}}{}% just name
    }%
}
\newcommand{\chref}[2][n]{%
    \ifthenelse{\boolean{onechapter}}{%
        \ifthenelse{\equal{#2}{ChPref}     }{\arxiv[Chapter ``\nameref*{ch:#2}'']{1801.09442}}{}%
        \ifthenelse{\equal{#2}{ChIntroPS}  }{\arxiv[Chapter ``\nameref*{ch:#2}'']{1801.09442}}{}%
        \ifthenelse{\equal{#2}{ChDetVal}   }{\arxiv[Chapter ``\nameref*{ch:#2}'']{2108.02706}}{}%
        \ifthenelse{\equal{#2}{ChStrucAli} }{\arxiv[Chapter ``\nameref*{ch:#2}'']{1801.09442}}{}%
        \ifthenelse{\equal{#2}{ChDBClass}  }{\arxiv[Chapter ``\nameref*{ch:#2}'']{1801.09442}}{}%
        \ifthenelse{\equal{#2}{ChFunc}     }{\arxiv[Chapter ``\nameref*{ch:#2}'']{1801.09442}}{}%
        \ifthenelse{\equal{#2}{ChIntroPred}}{\arxiv[Chapter ``\nameref*{ch:#2}'']{1712.00407}}{}%
        \ifthenelse{\equal{#2}{ChHomMod}   }{\arxiv[Chapter ``\nameref*{ch:#2}'']{1712.00425}}{}%
        \ifthenelse{\equal{#2}{ChSSPred}   }{\arxiv[Chapter ``\nameref*{ch:#2}'']{1801.09442}}{}%
        \ifthenelse{\equal{#2}{ChFuncPred} }{\arxiv[Chapter ``\nameref*{ch:#2}'']{1801.09442}}{}%
        \ifthenelse{\equal{#2}{ChIntroDyn} }{\arxiv[Chapter ``\nameref*{ch:#2}'']{1801.09442}}{}%
        \ifthenelse{\equal{#2}{ChThermo}   }{\arxiv[Chapter ``\nameref*{ch:#2}'']{1801.09442}}{}%
        \ifthenelse{\equal{#2}{ChMD}       }{\arxiv[Chapter ``\nameref*{ch:#2}'']{1801.09442}}{}%
        \ifthenelse{\equal{#2}{ChMC}       }{\arxiv[Chapter ``\nameref*{ch:#2}'']{1801.09442}}{}%
    }{
    \hyperref[ch:#2]{%
        \ifthenelse{\equal{#1}{n} }{Chapter \ref*{ch:#2}}{}% just number
        \ifthenelse{\equal{#1}{nn}}{Chapter \ref*{ch:#2} ``\nameref{ch:#2}''}{}% name & number
        \ifthenelse{\equal{#1}{N} }{``\nameref{ch:#2}''}{}% just name
      }%
  }%
}
\newcommand{\chrefname}[1]{\hyperref[ch:#1]{Chapter \ref*{ch:#1} ``\nameref{ch:#1}''}}
\newcommand{\partref}[1]{\hyperref[#1]{Part \ref*{#1}}}
\newcommand{\appref}[1]{\hyperref[app:#1]{Appendix \ref*{app:#1}}}

\newcommand{\figsource}[1]{\protect\footnote{Figure source location: \url{#1}}}

\newenvironment{penum}[1][\itshape i)\upshape]
{\begin{inparaenum}[#1]} {\end{inparaenum}}

\renewcommand{\arraystretch}{1.3}

\makeatletter \@beginparpenalty=5000 \makeatother

\newenvironment{bgreading}[1][]{
  \begin{mdframed}[%
      outerlinewidth=0,%
      linecolor=CornflowerBlue!30,%
      backgroundcolor=CornflowerBlue!30,%
      innerleftmargin=14,%
      innerrightmargin=14,%
    ]
	\ifthenelse{\equal{#1}{}}{}{
        \stepcounter{panel}
    	\subsection*{#1} 
    }
}{%
  \end{mdframed}
}

\usepackage{listings}
\definecolor{backcolour}{rgb}{0.95,0.95,0.92}
\definecolor{codegreen}{rgb}{0,0.6,0}
\definecolor{codegray}{rgb}{0.5,0.5,0.5}
\definecolor{codered}{rgb}{0.8,0,0.0}
\definecolor{codeblue}{rgb}{0.0,0,0.8}
\lstdefinestyle{codeStyle}{
    backgroundcolor=\color{backcolour},   
    commentstyle=\color{codegreen},
    keywordstyle=\color{codeblue},
    numberstyle=\tiny\color{codegray},
    stringstyle=\color{codegray},
    numbers=left,                    
    tabsize=2
} 
\newcommand{\sfrac}[2]{#1/#2}

\makeindex

\begin{document}

\setboolean{onechapter}{true}

\pagestyle{fancy}
\lhead[\small\thepage]{\small\sf\nouppercase\rightmark}
\rhead[\small\sf\nouppercase\leftmark]{\small\thepage}
\newcommand{\innerfoot}{\footnotesize{\sf{\copyright} Feenstra \& Abeln}, 2014-2023}
\newcommand{\outerfoot}{\footnotesize \sf Intro Prot Struc Bioinf}
\lfoot[\outerfoot]{\innerfoot}
\cfoot{}
\rfoot[\innerfoot]{\outerfoot}
\renewcommand{\footrulewidth}{\headrulewidth}

\mainmatter
\setcounter{chapter}{14}
\chapterauthor{\JvG*~\JvGid \and \MD~\MDid  \and \HM~\HMid \and \AR~\ARid \and \JV~\JVid \and \AF*~\AFid~~~\SA*~\SAid}

\chapterfootnote{* editorial responsability}

\chapterfigure{\includegraphics[width=0.5\linewidth]{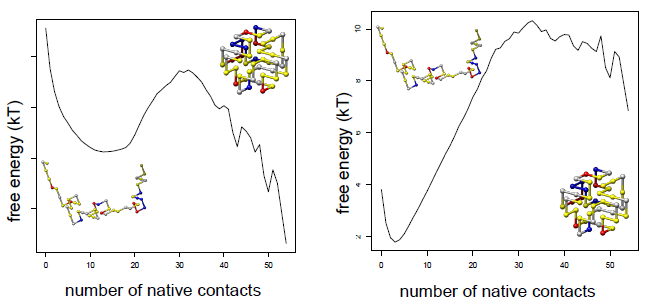}}
\chapter{Monte Carlo for Protein Structures}
\label{ch:ChMC}

\ifthenelse{\boolean{onechapter}}{\tableofcontents\newpage}{}

\section{Introduction}
In the previous chapter,  \chref{ChMD}, we have considered protein simulations from a dynamical point of view, using Newton's laws. In this Chapter, we first take a step back and return to the bare minimum needed to simulate proteins, and show that proteins may be simulated in a more simple fashion, using the partition function directly, as given in \chref{ChThermo}. 
We will assume basic knowledge on thermodynamics and statistical mechanics, as introduced there as well. It is particularly important to understand the relation between free energy and probability, in order to understand this chapter.
This means we do not have to calculate explicit forces, velocities, moments and do not even consider time explicitly. Instead, we heavily rely on the fact that for most systems we will want to simulate, the system is in a dynamic equilibrium; and that we want to find the most stable states in such systems by determining the relative stabilities between those states.

\section{Proteins in equilibrium}

Firstly, we will briefly revise our conceptual understanding of a dynamic equilibrium. In equilibrium, for each state in the system the number of particles moving into that state is equal to the number of particles moving from that state to a different state. 

\begin{figure}
 \centerline{
    \includegraphics[width=0.9\linewidth]{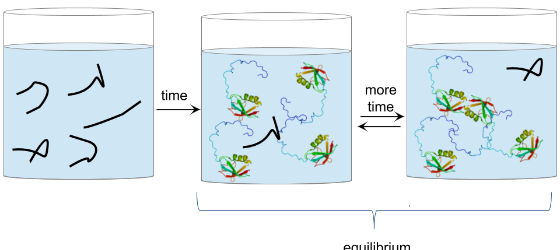}
}
  \caption{Proteins in equilibrium. Proteins are non-static entities. Over time, proteins constantly unfold and refold. When the proper folding of proteins is experimentally determined by for example by measuring the activity of the protein, the average behaviour over the ensemble of protein configurations in solution is determined rather than the behaviour of individual molecules. An equilibrium simulation of a single particle over time is equivalent to measurements on an ensemble multiple proteins in equilibrium - provided that they do not interact.}
  \label{fig:ChMC:Cartoons}
\end{figure}

Proteins in solution are dynamic systems, see \figref{ChMC:Cartoons}. Proteins constantly unfold and refold.  Once in equilibrium, the number of proteins moving from a folded to an unfolded state equals the number of proteins moving from an unfolded to a folded state, such that the fraction of folded and unfolded proteins will remain constant over time. We will see later in this chapter, that this also needs to hold for simulations in equilibrium; this is concept is called `detailed balance' (see \panelref{ChMC:detailed-balance} later in this chapter for more detail).

In this Chapter we will consider two systems: \begin{penum}
\item particles freely moving in a box; see \figref{ChThermo:marbles} in \chref{ChThermo}, and 
\item a simplified protein chain freely moving; 
\end{penum} 
see \figref{ChIntroDyn:Fold} in \chref{ChIntroDyn}.

In the first system, we consider the two macrostates: the colour separated and mixed states; here the positions of the particles define the specific configurations or microstates. In the second system we consider the folded and unfolded macrostate; here the positions of the particles (residues) in the chain define the specific configurations or microstates; for definitions of micro- and macrostates see \chref{ChThermo}

\section{The Purpose of Simulations}

Before we go into the technical details of simulations, we first reconsider what we typically want to learn from them. In Monte Carlo and Molecular Dynamics simulations, the main goal is to understand what the most stable state of the system is under certain conditions. For example, one can determine the stability of a certain fold, calculate the interaction strength of protein-protein or protein-ligand interactions, or the phase of the particles in the system under different conditions. If these interaction strengths are known, one can for example calculate the concentration needed for two proteins to start binding at a given temperature. In addition, determining the transition states between the most stable states in a system can recover mechanisms of function, when for example considering a binding or a folding process. 
The most stable state of a system is defined as the state with the highest probability and the lowest free energy. As discussed in \chref{ChThermo}, the free energy $F_A$ and probability $p_A$ of a macrostate $A$  are related as:

\begin{equation}
F_A = - k_B T \ln{(p_A)}
\label{eq:ChMC:Fstate}
\end{equation}
where $ {k_B} $ is the Boltzmann constant, $T$ is the temperature in Kelvin and $ {p_A} $ is the probability of state $A$. 

Moreover, as we previously discussed that the difference in free energy between two states calculated over a statistical ensemble approximates the difference in Gibbs free energy (i.e., $\Delta F_{A,B}  \approx \Delta G_{A,B}$), we also have:

\begin{equation}
\Delta G_{A,B} \approx - k_B T \ln{\frac{p_{A}}{p_{B}}}
\end{equation}

This means that once we have sampled the statistical ensemble of configurations appropriately, we can make a good estimate of the relative free energy between states.

\begin{figure}
  \centerline{
    \includegraphics[width=0.9\linewidth]{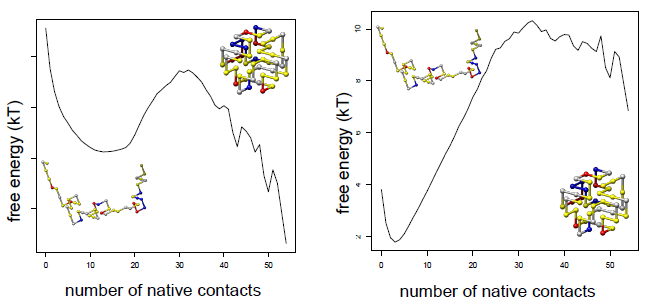}
  }
  \caption{Free energy of a protein in a 3D cubic lattice model of a protein at high and low temperature. Left: at low temperature, the system with the largest number of native contacts is the most stable. The low enthalpy has the largest influence on the free energy of the system, and therefore the configuration with the largest number of favourable interactions is the most stable. Right: at high temperature, the state with the largest entropy has the lowest free energy and is therefore more stable than the native state. }
  \label{fig:ChMC:freeEnergy}
\end{figure}

\figref{ChMC:freeEnergy}  illustrates the difference in  free energy between the folded and unfolded state at two different conditions. It is this relative free energy that determines the stability of the respective states. 

Note that, with any simulation technique, it is only possible to calculate relative free energies. If we wanted to get absolute free energies - we would need to calculate the full partition function, which is (computationally) intractable. 
Nevertheless, absolute free energies may be estimated from reference points for which the full partition function can be calculated. Such calculation go beyond the scope of this book, but are described in \citet{FrenkelSmit}.

\section{Comparison to experiments}

Similar to simulations, relative free energies between well defined states can be obtained from experiments. Differences in enthalpy ($\Delta H$) can also  be measured directly between states\cite[e.g.,][]{Kardos2004}.

With some experiments, we can obtain information about the configurational ensemble of proteins in solution. For example, Hydrogen-Deuterium exchange experiments can reveal the fraction of surface exposed residues of an ensemble in solution \cite{Englander2017}. In simulations, we can estimate such observables on the macrostate through ensemble averages, by averaging over the microstates:

\begin{equation}
\langle a \rangle = \frac{\sum_{i}{a_i p_i}}{\sum_{i}{p_i}}
\end{equation}
From a simulation, we can simply calculate an ensemble average $\langle a \rangle$, by averaging a certain property $a$ over all the sampled microstates (or configurations) $i$. See \chref{ChThermo} \secref{ChThermo:EnsAvg} for a more detailed explanation of ensemble averages.

\section{Monte Carlo Alogrithm}

The Monte Carlo algorithm can be used in simulations with a constant number of particles, volume and temperature, also referred to as NVT ensemble; see \chref{ChThermo} \secref{ChThermo:Ensembles}. In the Metropolis Monte Carlo algorithm one can sample the partition function directly, which means we do not need to consider forces, velocities or time. 
What we do need in order to sample the partition function, is a way to obtain the potential energy of specific configurations. 

\subsection{Potential energies}

We can calculate the potential energy $E_i$ for a micro state $i$, if we consider all pairwise interactions between the particles:

\begin{equation}
E_i = \frac{1}{2}\sum_{k=0}^{k=N} \sum_{l=0}^{l=N} \epsilon_{(k,l)} C_{(k,l)}
\label{eq:ChMC:Energy}
\end{equation}

Here $\epsilon_{(k,l)}$ are the pairwise interaction energies between particles $k$ and $l$, and $C_{(k,l)}$ indicates if the two particles interact  with each other, which would depend on the distance of the two particles.

We can also use continuous interaction potentials, such as the Lennard-Jones potential. In that case, the pairwise particle interaction energies ($\epsilon_{(k,l)}$) also depend on the distances between particles as shown in \figref{ChMD-lennard-jones} in \chref{ChThermo}.

\subsection{Sampling the partition function}

As explained in \chref{ChThermo}, the partition function $Z$ can be used to calculate the free energy and describe the state of the system (i.e the macrostate). From the Boltzmann distribution we have:

\begin{equation}
\label{eq:ChMC:Boltzmann}
p_i = \frac{e^{- \frac{E_i}{k_B T}}}{Z}
\end{equation}
where $Z = \sum_{i}{e^{- \frac{E_i}{k_B T}}}$.

If we know all the possible configurations (microstates) of the system, it is possible to calculate the absolute free energy landscape of the system from \eqref{ChThermo:FETS2} in \chref{ChThermo}. Note  that for a continuous three-dimensional system ($\mathbb{R} ^3$) with a constant finite number of particles the partition function becomes an integral over the full three-dimensional space, rather than a sum over all possible configurations. 

However, in a simulation, computation of the full partition function is intractable. Instead, we aim to sample those configurations (microstates) with the largest contribution to the total free energy; from \eqref{ChMC:Boltzmann} we can see that the microstates with the highest probabilities are the microstates with low energies. However the contribution low energy microstate become smaller at high temperatures.

\subsection{The Metropolis Monte Carlo algorithm}

\begin{figure}[!ht]
  \centerline{
    \includegraphics[width=0.5\linewidth]{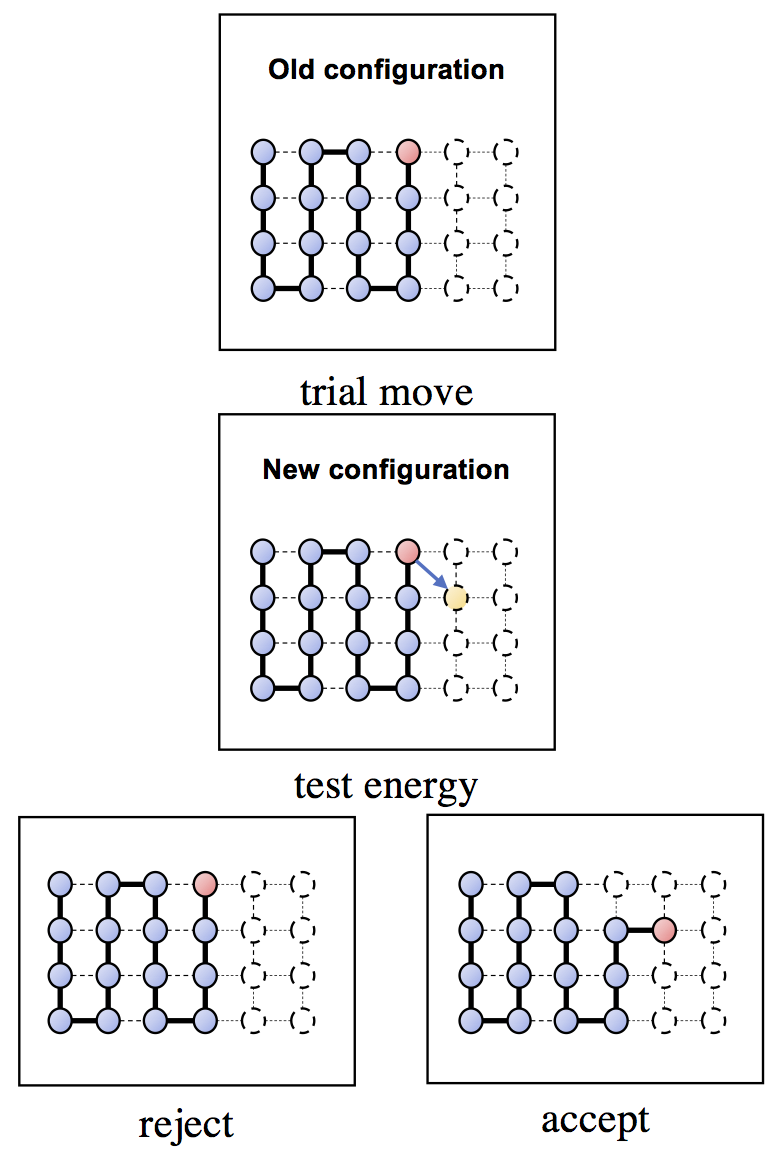}
  }
  \caption{Trial move in a Monte Carlo simulation. Based on whether the change in energy of a random configurational change is favourable or not, it will be either accepted or rejected as the new state of the system. Unfavourable moves are accepted with a probability equal to the Boltzmann factor. Here a coarse-grained model of a protein on a 2D square lattice is shown to exemplify the algorithm.}
  \label{fig:ChMC:monteCarlo}
\end{figure}

The Monte Carlo algorithm is a stochastic algorithm that only depends on the potential energy of the system. The temperature, volume and number of particles in the system are kept constant. Additionally, the algorithm assumes the system is in equilibrium. 

The key idea in the Monte Carlo algorithm is to make sure the probabilities of the sampled (micro)states follow the Boltzmann distribution. 
This can be achieved in a simple manner: by generating a random move, and consistent rule - the Boltzmann acceptance criterion.

In the algorithm random moves are proposed to change the configuration of the system: randomly chosen particles are moved by a random, but typically small, displacement, as shown in \figref{ChMC:monteCarlo}. Now we have two configurations, the `old' configuration and the `new' configuration. For both configurations we can calculate an explicit potential energy, using \eqref{ChMC:Energy}. These energies of the microstates can be used to calculate the Boltzmann factor $B$:
\begin{equation}
B = e^{- \frac{E_\text{new} - E_\text{old}}{k_B T}}
\label{eq:ChMC:Boltz}
\end{equation}
where ${E_\text{new}}$ is the energy of the new state, $E_\text{old}$ is the energy of the previous state, ${k_B}$ is the Boltzmann constant and {T} is the temperature.

When the new configuration has a lower energy than the old configuration, i.e.,\@ $E_\text{new} <= E_\text{old}$ we always accept the move, note that in that case $B>1$. If, on the other hand $E_\text{new} > E_\text{old}$ we use the Boltzmann factor and a random variable $r \in [0,1]$ to determine if the move will be accepted: the move will only be accepted if $r < B$.

Note that in the latter case, the system will actually get a more unfavourable energy after the move. At high temperatures, the Boltzmann factor will be close to one even if the energy difference between the old and new state is large; hence, at high temperatures the majority of moves will be accepted. This will lead to the enthalpic contribution becoming less dominant. This can be directly compared to the classical thermodynamics relation $\Delta G = \Delta E - T\Delta S$, which states that the entropy becomes more dominant at higher temperatures. The full MC algorithm is listed in \figref{ChMC:MCpseudo}.

\begin{figure}

\begin{lstlisting}[language=Python]
# num_cycles: how many cycles of random sampling
# N: number of particles (or residues)
# V: volume
# T: the temperature 
# C: initial configuration of the particles (protein)
def monte_carlo(num_cycles,N,V,T,C):
    config_old = C
    for x in range(num_cycles):
        # pick a particle (residue) to displace 
        # randint() is random integer generator
        x = randint(0, len(N)-1)
        # move the chain by generating 
        # a new configuration for particle x
        # note that the new configuration is generated
        # within a constant volume (V)
        config_new = generate_config(config_old,x,V)
        # calculate the old and new interaction energies 
        # for particle x
        E_new = Energy(config_new)
        E_old = Energy(config_old)	     
        # calculate Boltzmann factor, given kT
        boltz = exp(-(E_new - E_old)/k*T)     
        # acceptance criterion:
        acc = min(1.0, boltz))
        # rand() gives random number between 0 and 1
        # accept move if rand() is smaller than the
        # acceptance criterion
        if(rand() < acc):
            # move is accepted	
            config_old = config_new
            system_Energy += (E_new - E_old)
        #end if
        #sample at every step, to calculate p_i
        sample(config_old)
    # end for loop
# end Monte Carlo
\end{lstlisting}
\caption{Monte Carlo algorithm for molecular simulations in pseudo code Python style.}
\label{fig:ChMC:MCpseudo}
\end{figure}

To obtain a correct sampling of the partition function, sampling needs to be performed after every move, regardless of whether it is accepted or rejected; this means that for a rejected move, we count (sample) the old configuration again (!). Note that this may make more intuitive sense if you consider a state that is already close to the free energy minimum (e.g., a folded state, and try to move away from this state (e.g., partially unfold the protein), which may be rejected in most trial moves. In this case, the low free energy state (e.g., folded state) will be sampled very often - but only if we also sample the old configuration after a rejected move.

From the simulation, the probability for a particular macrostate can be determined by calculating the fraction of configurations within the state, and those sampled outside of this state. Subsequently, the relative free energy of that state can be calculated using \eqref{ChMC:Fstate}.

As the simulation should be in equilibrium, in theory the  starting state of the system should not matter. In practice, it is wise to check if there is indeed no flux during the simulation: if the simulation starts from a high free energy (unlikely) state, it may get stuck in a local minimum for a while, effectively not sampling the partition function evenly.

\begin{bgreading}[Detailed balance]
\label{panel:ChMC:detailed-balance}

Detailed balance is a way of making sure equilibrium is kept in a Monte Carlo simulation. In other words, it ensures there is no net flux between states over time.
 Hence the number of accepted moves from a state S1 to state S2 needs to equal the number of accepted moves from the state S2 into that state S1, for any two states S1 and S2 in the system. This can be expressed as follows:
\begin{equation}
{N_{S1} * P_{acc} \left(S1, S2 \right) } = {N_{S2} * P_{acc} \left(S2, S1 \right)}
\label{eq:ChMC:db}
\end{equation}
Here $ {N_{S1}} $ and $ {N_{S2}} $ represent the number of times states A and B are visited, respectively, and $ { P_{acc} \left(i, j \right)} $ is the probability that the move from state $i$ to state $j$ is accepted.  
One can show that the Boltzmann acceptance criterion used in the Monte Carlo algorithm, $P_{acc} \left(S1, S2 \right) = \min \left({ e^{- \frac{E_j - E_i}{k_B T}}, 1 }\right)$ adheres to this rule. 

Note that $N_{S1}$ and $N_{S2}$ can be replaced by the probabilities that the states are visited, i.e., $p_i$ and $p_o$, respectively.

\centerline{
    \includegraphics[width=0.95\linewidth]{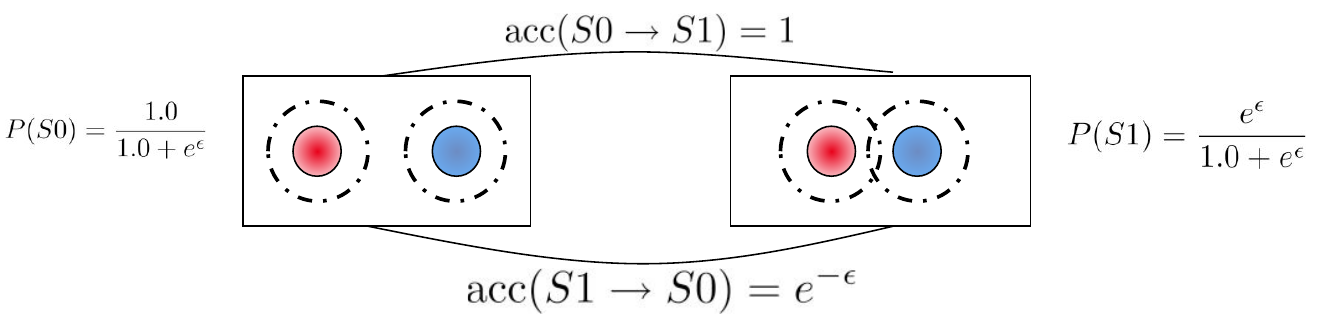}
  }

Here, we will simply demonstrate that the Monte Carlo acceptance criterion satisfies detailed balance with the simple example shown above. We consider a system with two particles in solution and only two possible states: either they are separated and do not interact (left) or they are bound and have a favourable interaction, with an interaction energy $-\epsilon$ (right). In this case we have two states: $S0$ (separated) and $S1$ (bound), hence ${E_0 = 0}$ and ${E_1} = {-\epsilon}$. For simplicity, we can set ${k_B T} = 1$. Using the  probabilities from \eqref{ChMC:Boltzmann} , we get ${p_0} = {\sfrac{e^{0}}{(e^{0} + e^{\epsilon})}}$,  ${p_1} = {\sfrac{e^{\epsilon}}{(e^{0} + e^{\epsilon})}}$,  ${P_\text{acc} {(S0 \rightarrow S1)} = 1}$ and ${P_\text{acc}(S1 \rightarrow S0)} = e^{-\epsilon}$. Substituting this into \eqref{ChMC:db} gives:
\begin{equation}
{ {\frac{e^{0}}{e^{0} + e^{\epsilon}}} * 1} = {{\frac{e^{ \epsilon}}{e^{0} + e^{ \epsilon}}}* e^{- \epsilon}}
\end{equation}
Since ${e^0} = {e^\epsilon * e^{-\epsilon}} = 1$, the left and right hand side of the equation are equal. Therefore, the system is in equilibrium and detailed balance is satisfied.

For Monte Carlo simulations it is essential that detailed balance is kept, else the results of the simulation will be non-physical as the partition function will not be sampled correctly.
Note that there are many ways to break detailed balance, for example by not sampling after rejected moves.
\end{bgreading}

\section{Applications of Monte Carlo for proteins}
\subsection{A simple protein lattice model}
\label{sec:ChMC:lattice}

Full-atomistic simulations are computationally very demanding; in fact so demanding that it is still computationally too expensive to simulate the folding of proteins or realistic size ($\sim$100 residues) that form fully hydrophobic cores, as explained at length in \chref{ChMD}. Therefore, it is very useful to simplify such a system into a lattice model \cite{Sali1994,Coluzza2003,coluzza04,Abeln2008,Abeln2011,Abeln2014,vanDijk2016}. 
The residues are placed onto a regular cubic-lattice, which means we have a discrete rather than a continuous three dimensional space. This greatly reduces the number of possible configurations for the protein chain. Nevertheless, for real size proteins the number of possible configurations is still computationally intractable, even on a discrete lattice.

\begin{figure}[h]
  \centerline{
    \includegraphics[width=\linewidth]{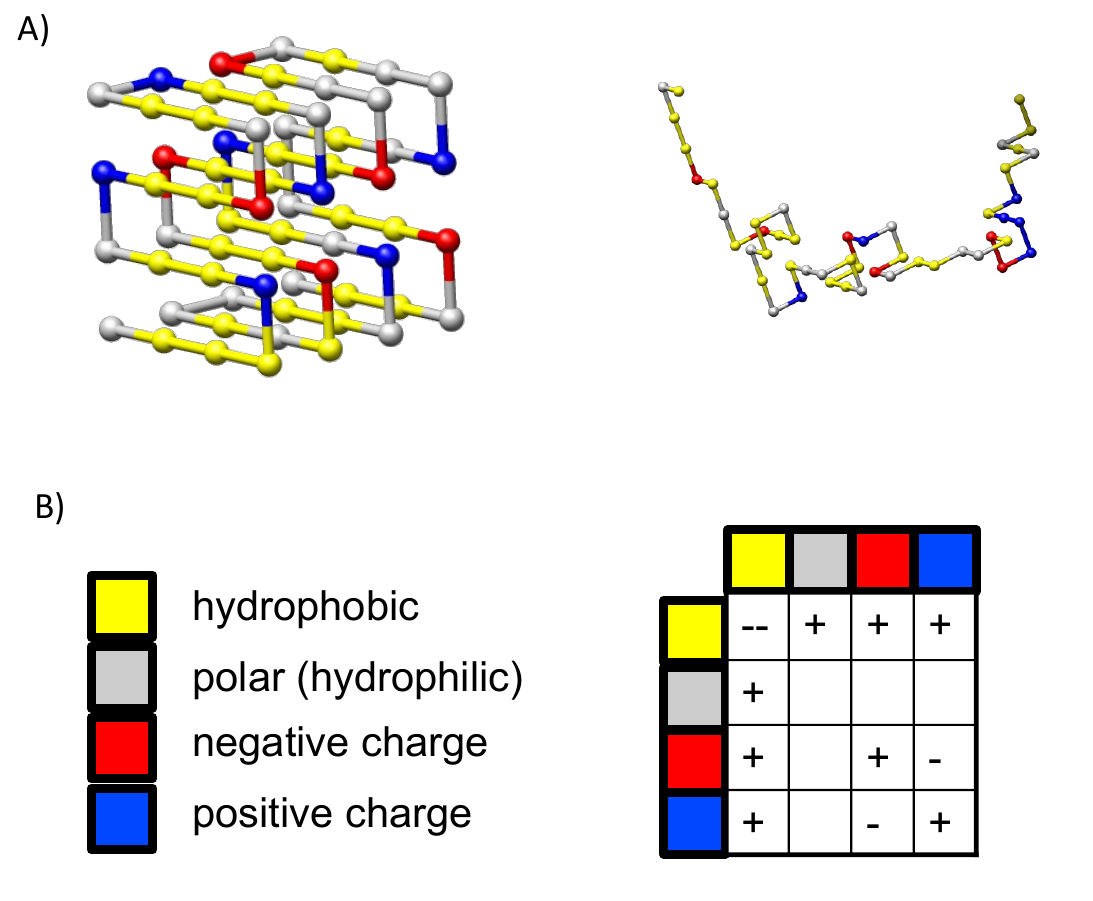}
  }
  \caption{ Simple 3D lattice model of a protein. A) a folded and unfolded configuration on the cubic lattice. The residues in the protein are placed on a 3D grid. Note that on the cubic lattice a residue has a maximum of four contacts with other residues - this is relatively similar for the average contact number of residues in real proteins. 
  B) Schematic interaction energies.  For simplicity, the amino acid pair potential is schematically shown in terms of interaction energies  ($\epsilon_{(k,l)}$) for Hydrophobic residues indicated in yellow, polar residues in grey, positively charged residues in red and negatively charged residues in blue. 
  }
  \label{fig:ChMC:latticeProperties}
\end{figure}

\begin{figure}[h]
  \centerline{
    \includegraphics[width=0.9\linewidth]{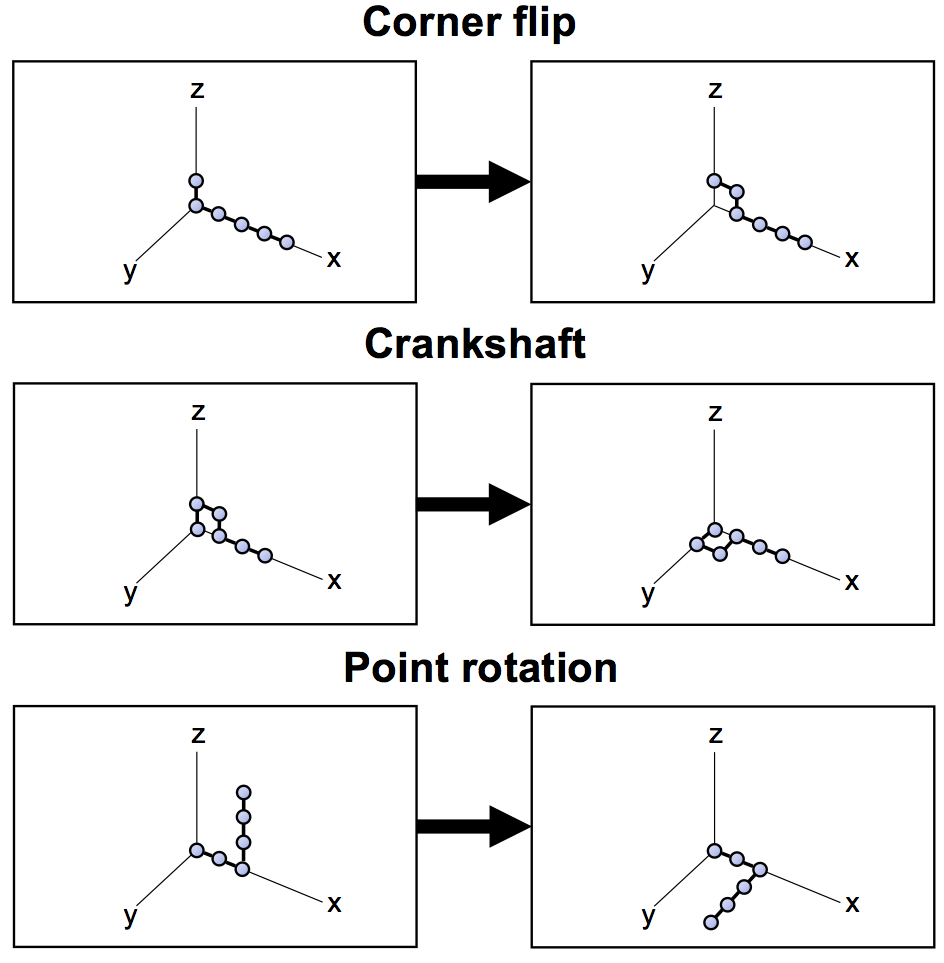}
  }
  \caption{Moves on a cubic lattice. Three different moves on a cubic lattice are shown: the corner flip, crankshaft and point rotation. Each of the moves ensure the chain is not broken after the move. In order to keep detailed balance the reverse move needs to be equally probable as the forward move.}
  \label{fig:ChMC:latticeMoves}
\end{figure}

\figref{ChMC:latticeProperties} shows an example of a 3D lattice model. Two residues are considered in contact when the are on neighbouring positions on the lattice but are not linked with a peptide bond. Using this criterion, all pairwise interactions can be determined using
\begin{equation}
C_{k,l} = \begin{dcases*}   
        1 & \text{if k and l are in contact}\\   
        0 & \text{otherwise}\\   
\end{dcases*}
\label{eq:contactMC} 
\end{equation}
The strength of the interactions are defined in the matrix in \figref{ChMC:latticeProperties}. Now we can calculate the full potential energy over a specific configuration using \eqref{ChMC:Energy}. The model can be simulated with a Monte Carlo algorithm as shown in \figref{ChMC:MCpseudo}. To generate a new configuration, we should only consider moves, that are feasible on the cubic lattice. A set of possible moves, that do not break the chain, on a cubic lattic, are shown in \figref{ChMC:latticeMoves}.

\begin{figure}
  \centerline{
    \includegraphics[width=0.9\linewidth]{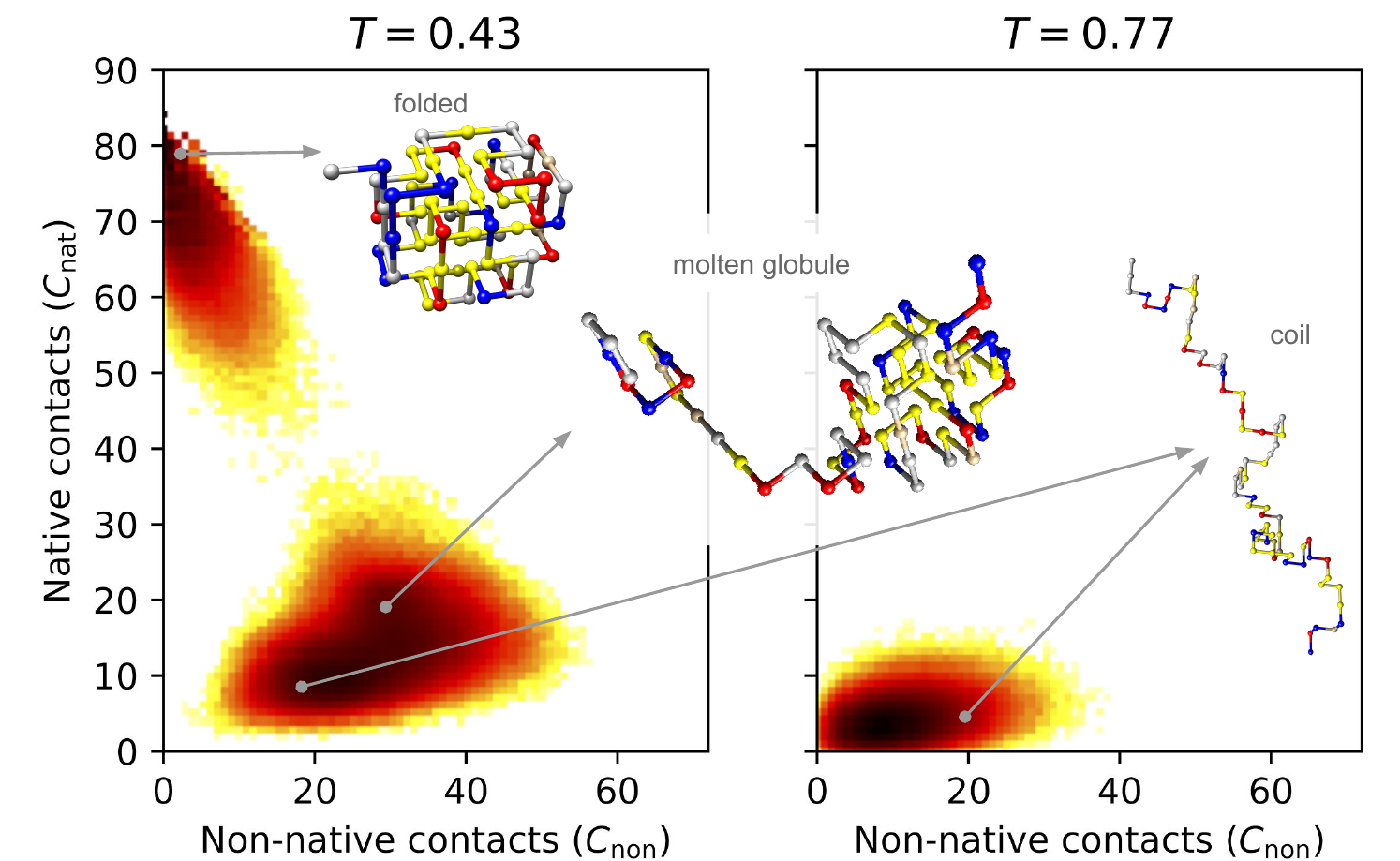}
  }

  \caption{Free energy landscape as a function of the number of native and non-native contacts in a lattice model, with the free energy values shown as heatmap colors (dark red is very low free energy; white is high free energy). At a high number of native contacts, the protein is in its native folded state (top left in the plots). At intermediate values of native and non-native contacts, the protein is in a molten globule state. At very low numbers of native and non-native contacts, the protein is an unfolded, coil-like state. The figure shows that at a low temperature (left), the free energy is low for the folded state, the molten-globule state and the unfolded state. At even lower temperatures (not shown here), both the molten globule state and the unfolded state become unstable. At high temperature (right) the free energy is lowest when there are very few native and non-native contacts in the protein, indicating that the unfolded state is the most stable.}
  \label{fig:ChMC:latticeMG}
\end{figure}

Monte Carlo and lattice models can be used to determine the most stable states of a protein under different physiological conditions. \citet{Dijkstra2018}, applied a Monte Carlo algorithm to a 3D protein lattice model to study the stability of a protein at different temperatures. Due to the simplified model, it becomes possible to obtain very extensive sampling of the conformational landscape, and allows details of the free energy landscape to be mapped out, as shown in \figref{ChMC:latticeMG}. The model describes three main states: the native folded state, molten globule state, and unfolded state. As shown in \figref{ChMC:latticeMG}, the native, molten globule and folded states  are all present at lower temperatures, whereas at high temperatures only the unfolded state has a low free energy.

Using such simulations, we can observe behaviour that is very similar to proteins in experimental settings: at high temperatures proteins unfold, due to the chain entropy. In this particular work, it was shown that proteins with the same fold, but with a different sequence, could have very different folding pathways and different intermediate molten-globule like states.

\subsection{Other applications in bioinformatics}
\label{sec:ChMC:mc_app_bioinf}

Fragment based structure prediction methods typically use Monte Carlo simulations to assemble decoy structures from the structural fragments \cite{Song2013}, see also \chref{ChIntroPred}. Here, Monte Carlo sampling is used as a search and optimisation technique. The simulation starts at a medium to high temperature, which is decreased step-wise throughout the procedure until $T=0$ and an energy minimum is reached. This process is called `simulated annealing'. 

Simulated annealing is also used in homology model building by MODELLER \cite{Sali1993}, and in proposing moves via Molecular Dynamics.  Here, the goal is to  optimise a configuration that adheres to structural constraints from a template structure, see also \chref{ChHomMod}. 

It is important to note that such optimisation procedures are fundamentally different from molecular simulation approaches that try to sample the partition function. In simulated annealing  only the (potential) energy is minimised, and not the free energy. In other words, entropy is not considered in the simulated annealing derived predictions. Moreover, typically non-physical energies are included in the energy function, such as distant constraints on specific residues. It is important to realise that we cannot use such optimisation techniques to consider folding or binding mechanisms.

\begin{bgreading}[Hybrid MC \& MD simulations]
\label{panel:ChMC:hybrid-MC-MD}

Proteins are very long molecules (polymers, or polypeptides). This means that any moves along the chain are generally correlated: neighbouring atoms in the chain cannot move independently from each other.
This means that Monte Carlo moves on single atoms or residues -- in case of coarse grained models -- can be very inefficient. One way of overcoming this is to generate collective moves; in structure prediction the fragment based approach of Rosetta \cite{Song2013} is very efficient. For molecular simulations, often a hybrid approach gives extremely efficient sampling  \cite{Woo2004,Pool2012,Yang2016}: here, the smaller moves are implemented as a series of MD steps. These trajectories may then be rejected or accepted according to the rules based on the Boltzmann factor, making the higher level moves stochastic. The advantage of such an approach is that a multitude of enhanced sampling techniques can easily be applied within a high level MC simulation, using low level MD moves and a force field parametrised for MD. In such hybrid simulations, time development and (hydro)dynamics are not conserved. 
\end{bgreading}

\section{Enhanced sampling techniques}
\label{sec:ChMC:enhanced-sampling}
As explained in previous sections, the relative free energy of a state can be calculated from the fraction of time spent in that state during a simulation. Low free energy states correspond to a high probability of sampling. This means that during a simulation, mainly the most stable states are sampled. On the other hand, sampling of high energy states is much more difficult: in severe cases, there may be no sampling of such states all together. This is particularly troublesome if these higher energy states lie in between two stable states, since such states form a `barrier' between two stable states.
An example of this was already shown in \figref{ChIntroDyn-FreeEnergy}. %
In order to calculate the relative free energy of the two stable states, it is essential to also sample the path connecting them. There are different tricks that can be applied to improve sampling in these regions and obtain a free energy landscape over the entire region of an order parameter.

Here, we will discuss two methods for enhanced sampling: Umbrella Sampling and Replica Exchange/Parallel Tempering. Both methods can be applied within MD simulations as well as MC, but are more easily implemented in MC. Moreover, the exchange steps in Replica Exchange are essentially Monte Carlo moves.

\subsection{Umbrella Sampling in MC}

Umbrella Sampling is one of the more simple enhanced sampling techniques. In Umbrella sampling, a value of the order parameter (e.g.,\@ the distance between two interacting proteins) is chosen around which one wants to sample. 

In a Monte Carlo simulation this is extremely easy to implement. The only thing we need is a good order parameter. If for an order parameter $x$ we want to sample a barrier region between $a$ and $b$, we need to ensure that the path sampled by the MC algorithm rejects any steps going to a microstate where $x<a$ or $x>b$.  Remember that the (sampling) probability of a state has a direct relation with the free energy of that state: from \eqref{ChTermo:pa_pb} in \chref{ChThermo} we can derive $p_A = e^{F_A / k_B T}$, where $F_A$ is the free energy relative to the other sampled states.  Now, we can easily understand that the sampling probability of a state will go up, if the system is not allowed to visit the low free energy states of the system. In other words, if we choose the interval  between $a$ and $b$ to be small enough, such that sampling is focused on the high free energy states only, the probability of sampling the barrier goes up. Now we can split the entire free energy landscape in multiple intervals. For each interval, we can approximate a free energy curve, which can be stitched together in a final step. Generally, the steeper the slope of the free energy curve with respect to the order parameter, the more intervals we need. Once we have all the free energy curves for the the intervals, we need to stitch them together, this can be done by curve fitting; this will work much better, if there is overlap between the intervals. For more details, please see \citet{FrenkelSmit}

\subsubsection{Umbrella sampling using quadratic potentials}

In MD simulations, we cannot simply reject moves or add a "hard wall". Instead, an artificial energy penalty is added around a selected point, such that it becomes very unfavourable for a protein to deviate far from this point. This penalty is called the `Umbrella potential' ($E_{Umbrella}$) and takes the form of a quadratic equation:
\begin{equation}
E_{umbrella} = k_{umbrella} \left( d - d_0 \right)^2
\end{equation}
where a higher value of $k_{umbrella}$ indicates a steeper penalty for deviating distance ${d - d_0}$ from the selected point $d_0$. 

\begin{figure}
  \centerline{
    \includegraphics[width=0.9\linewidth]{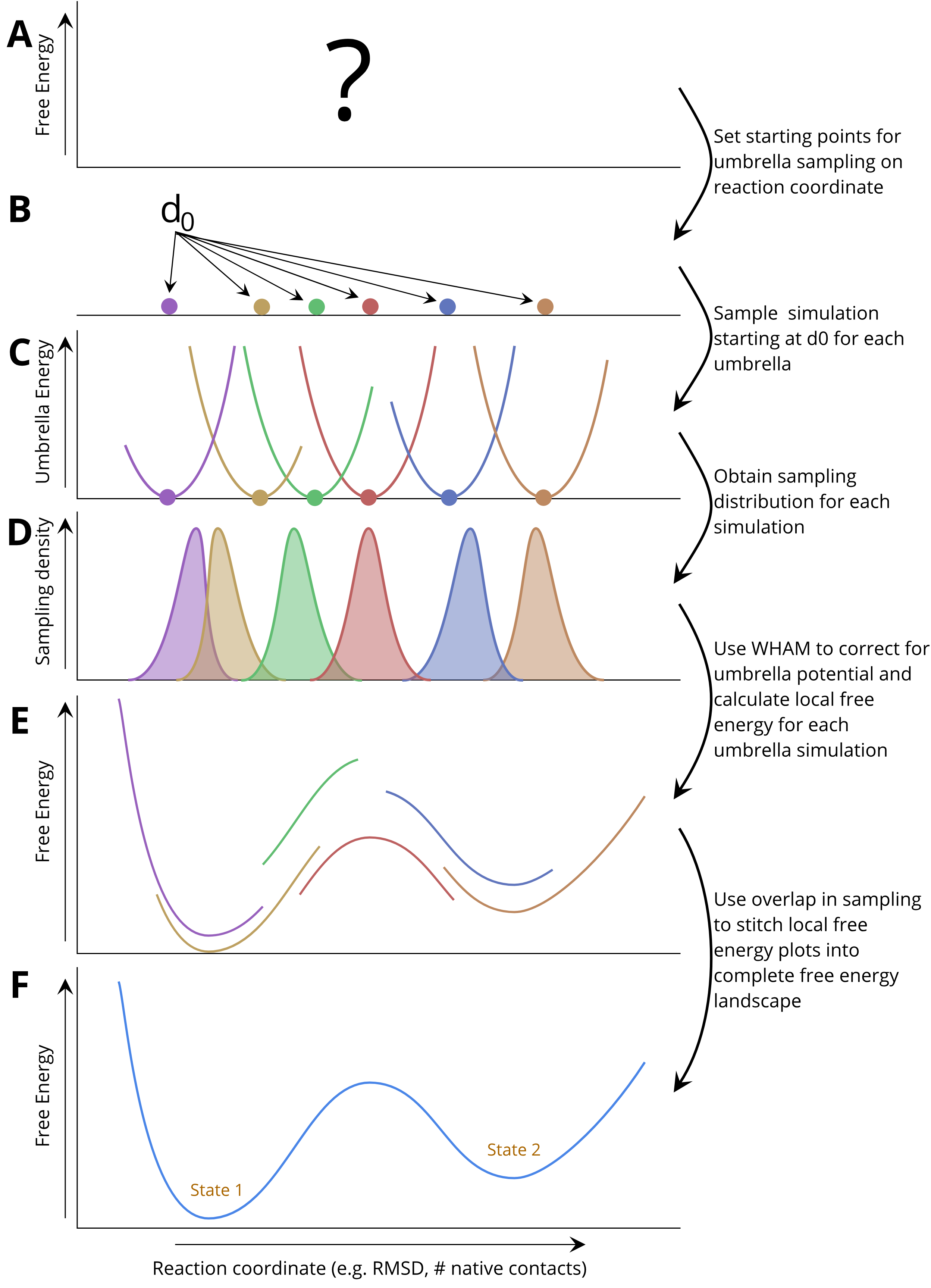}
  }
  \caption{Schematic overview of an umbrella sampling for an MD simulation (see main text for further details). (A) Choice of the reaction coordinate (RC). (B) Apply umbrella potentials on selected values of the RC. $d_0$ is the minimum of the umbrella in terms of the RC (C) Individual sampling around selected coordinates. (D) Density of sampling along the RC for each simulation. (E) Using weighted histogram analysis method (WHAM). (F) Joining the local free energy landscapes into a complete free energy landscape around the RC. Note that the sampling overlap is essential to create the final free energy landscape.}

  \label{fig:ChMC:umbrellaLandscape}
\end{figure}

Now we can draw such umbrellas over the entire range of the order parameter of interest, as shown in \figref{ChMC:umbrellaLandscape}. The name `Umbrella sampling' originates from the shape of the penalty curve.

Finally, to obtain the true free energy landscape from the different simulations, the obtained energies need to be corrected for the umbrella potential. This can be done using
\begin{equation}
\langle A \rangle = {\frac{\langle \frac{A}{w} \rangle_w}{\langle \frac{1}{w} \rangle_w }}
\end{equation}
where $A$ is the property of interest and $w$ the weights of the sampling. The correctness of the final fit, stitching the intervals together to obtain a free energy curve can be improved using the WHAM  method \cite{Grossfield2003}.

\subsubsection{Umbrella sampling procedure}
The umbrella sampling procedure can be summarised as follows (\figref{ChMC:umbrellaLandscape}):
\begin{itemize}
    \item[A] First, a reaction coordinate needs to be defined and an estimate of a range of values of this reaction coordinates that captures the relevant protein dynamics needs to be made.
    \item[B] Subsequently, multiple points within this range of the reaction coordinate are chosen to initiate the simulations. On each of these starting points, an umbrella potential is applied that adds an energy penalty to the simulation whenever the value of the reaction coordinate deviates from the starting point. The penalty is zero at the starting point, and increases quadratically as the distance from this value of the reaction coordinate increases ($E_{umbrella} = k_{umbrella}\left( d - d_0 \right)^2$), though other functions for the energy penalty may be chosen as well.
    \item[C] While running the simulations, at each point the value of the reaction coordinate and corresponding umbrella energy is sampled.
    \item[D] After the simulations are completed, the density of sampling along the reaction coordinate is calculated for each simulation.
    \item[E] Using weighted histogram analysis method (WHAM), the free energy profile can be corrected for the added umbrella potential and a local free energy landscape can be created.
    \item[F] Using the overlap in the regions of the reaction coordinate that were sampled between simulations, the local free energy landscapes can be stitched together into a complete free energy diagram of the sampled region of the reaction coordinate. Note that the sampling overlap between simulations is necessary to be able to create the final free energy landscape. If overlap is insufficient or lacking in any area, additional simulations need to be run initiated in this area to obtain a higher sampling density.
\end{itemize}



\begin{bgreading}[Replica Exchange or Parallel tempering]

Parallel tempering, also known as temperature replica exchange, is another enhanced sampling technique. The key idea is that some transitions may be more easily sampled at different, typically higher, temperatures than the temperature of interest.

This approach consists of letting a number simulation boxes run simultaneously, while each box visits different temperatures during the parallel tempering procedure. These simulations are referred to as replicas, that can run in parallel. At fixed time intervals (MD) or number of steps (MC), attempts are made to exchange temperatures between the different simulation boxes. Attempting to exchange temperatures between replicate simulations follows a Monte Carlo procedure, which is best described as performing a Monte Carlo move in temperature space. 

With this Monte Carlo move, we need to ensure that detailed balance is observed, such that we have equal probabilities for the forward and backward swaps. It can be shown that the following rule for accepting moves, indeed keeps detailed balance.

\begin{equation}
P_{acc}(S1 \rightarrow S2) = \min(1, e^{(\beta_1-\beta_2)(E_1-E_2)})
\label{eq:PT}
\end{equation}

Here the variable $\beta_i = \frac{1}{kT_i}$. Note that $\beta$ is often used instead of T, to make manipulation of equations in thermodynamics easier. $E_i$ are the potential energies of the states to be swapped. A formal proof to show this acceptance rule adheres to detailed balance, which can be found in more details in\cite{FrenkelSmit}.

A little care needs to be taken, how the temperatures of the different simulation replicas are chosen. It is important that the temperatures are swapped sufficiently.
As a rule of thumb, one accepted exchange out of three trials is considered reasonable. 
A replica exchange procedure can be considered to be finished if all replica boxes have visited all temperatures several times. Then, the system has heated up and cooled down several times. If swaps between specific temperatures do not occur during the procedure, this suggests that these temperature may lie close to a transition point, and typically the interval between temperatures need to be made smaller, to allow for sufficient sampling.

\end{bgreading}

\section{Monte Carlo vs. Molecular Dynamics}

Now we have considered two simulation protocols, Molecular Dynamics (MD) and Monte Carlo (MC), both can be used to study the same properties of a system, namely the stability of states and the transitions between them. Using either technique, the free energy landscape can be calculated along a chosen order parameter (or multiple order parameters). However, in practice it is not possible to sample a complete folding pathway of a real-size protein in a full-atomistic model with either of the two techniques. Thus, we cannot exhaustively cover the whole free energy landscape, and we typically refer to the simulation process as \emph{sampling} states in the free energy landscape. Both techniques should maintain detailed balance, and sample the Boltzmann distribution. A short summary of main differences is provided in \tabref{ChThermo:MCvsMD}.

MC is an intrinsically stochastic method that depends on random moves to determine a simulation path. To calculate the next state of a system, only the energy difference between the old and the new state needs to be known. Any forces, velocities, momenta, and time are ignored in MC. This large simplification of the system makes MC simulations much faster to execute and much easier to code than MD.

MC simulations natively sample an NVT ensemble, while MD on the other hand natively samples an NVE ensemble, see also \citet{FrenkelSmit} for more details.

MD is theoretically a deterministic simulation,  however, in practice, due to limits in computational precision, and the use of a thermostat and/or barostat, MD is it is not deterministic.

Most biological systems are naturally exposed to an environment with constant temperature, i.e.,\@ they exist within larger systems with constant exchange of heat between the system and its surroundings, leading to a constant temperature of the considered system. Therefore, NVT is often a more natural choice. This means that for most practical cases we will need a thermostat in MD simulations; this (re)tunes the velocities of particles in such a way that the temperature is kept constant throughout the simulation.

Since MD captures dynamics explicitly, it is possible to include effects such as hydrodynamics (e.g.,\@ movements of water in direct vicinity to a moving part of the protein). In MC, because the forces, speeds, and momenta of all the particles are not known, collective moves, incorporating multiple particles, often need to be added explicitly to speed up the simulation.

Lastly, due to the simplicity of the MC algorithm, it is much more straightforward to implement enhanced sampling techniques (see section below) in an MC simulation. If we want to consider large systems, such as proteins that (re)fold, enhanced sampling techniques are essential to allow even sampling within a range of the order parameter during the simulation. 

\begin{table}
\begin{center}
\begin{tabular}{%
        >{\raggedright\arraybackslash}p{0.23\linewidth}%
        >{\raggedright\arraybackslash}p{0.35\linewidth}%
        >{\raggedright\arraybackslash}p{0.35\linewidth}%
        }
\hline
&\bf MC & \bf MD \\
\hline
\hline
algorithm &stochastic & deterministic \\
native ensemble &NVT  & NVE \\
advantages & easier to code & explicit dynamics \\
           & easier to implement enhanced sampling & time development\\
disadvantages & need collective moves for efficient sampling & need integrable forces \\
            & fewer simulation packages available & thermostat required for NVT \\
\hline
\end{tabular}
\caption{Monte Carlo (MC) versus Molecular Dynamics (MD) simulations.}
\label{tab:ChThermo:MCvsMD}
\end{center}
\end{table}

\section{Key points}

\begin{compactitem}
\item When a system is in equilibrium we do not have to simulate velocities and time explicitly in order to obtain relative free energies
\item Monte Carlo samples the partition function of systems in equilibrium
\item Monte Carlo is a stochastic sampling method
\item It is straightforward to use enhanced sampling techniques in the Monte Carlo framework
\item Molecular simulations need to keep detailed balance in order to adhere to statistical mechanics
\item In structural Bioinformatics many ideas of molecular simulation are used, sometimes with shortcuts that mean the sampled ensembles may be non-physical.
\end{compactitem}

\section{Further reading} 
\begin{compactitem}
\item \citet{Vlugt2008}
\item \citet{FrenkelSmit}
\end{compactitem}

\section*{Author contributions}
{\renewcommand{\arraystretch}{1}
\begin{tabular}{@{}ll}
\ACtxt: &   JvG, MD, HM, AF, JV, SA \\
\ACfig: &   JvG, MD, AR, AF, SA, \\
\ACref: &   JvG, JV, AF, SA\\
\ACproof:&  AF, JV, HM, SA \\
\ACfb:  &   AR \\
\ACeds: &  JvG, SA
\end{tabular}}

\mychapbib
\clearpage

\cleardoublepage

\end{document}